\documentclass[12pt]{article}
\usepackage{latexsym,amssymb}
\usepackage{amsfonts}

\catcode`\@=11

\def\section{\@startsection{section}{1}{\z@}{3.5ex plus 1ex minus
 .2ex}{2.3ex plus .2ex}{\bf}}

\def\thesubsection{\arabic{section}.\arabic{subsection}}
\renewcommand{\subsection}[1]{\addtocounter{subsection}{1}
\vspace{2.5mm}\par\noindent {\it \thesubsection . #1}\par
 \vspace{0.5mm} }
\catcode`\@=12
\newfont{\mbm}{msbm10 scaled\magstep1}

\DeclareFontFamily{U}{rsf}{} \DeclareFontShape{U}{rsf}{m}{n}{
  <5> <6> rsfs5 <7> <8> <9> rsfs7 <10-> rsfs10}{}
\DeclareMathAlphabet\Scr{U}{rsf}{m}{n}

\mathchardef\varGamma="0100 \mathchardef\varDelta="0101
\mathchardef\varTheta="0102 \mathchardef\varLambda="0103
\mathchardef\varXi="0104 \mathchardef\varPi="0105
\mathchardef\varSigma="0106 \mathchardef\varUpsilon="0107
\mathchardef\varPhi="0108 \mathchardef\varPsi="0109
\mathchardef\varOmega="010A

\def\drawbox#1#2{\hrule height#2pt\hbox{\vrule width#2pt height#1pt
 \kern#1pt\vrule width#2pt}\hrule height#2pt}

\def\Asym#1#2{\vcenter{\vbox{\drawbox{#1}{#2}\kern-#2pt\drawbox{#1}{#2}}}}

\begin{document}
\begin{titlepage}

\thispagestyle{empty}

\begin{flushright}
\hfill{CERN-PH-TH/2004-186}
\end{flushright}

\vspace{35pt}

\begin{center}{ \LARGE{\bf
No-scale supergravity from higher dimensions \footnote{To appear
in the proceedings of Strings 2004, June 28-July 2 Paris. }}}

\vspace{60pt}

{\bf R. D'Auria $^\star$, S. Ferrara $^{\dagger\ddag\sharp}$ and
M. Trigiante $^\star $}

\vspace{15pt}

$\dagger${\it  CERN, Physics Department, CH 1211 Geneva 23,
Switzerland. }\\ $\ddag${\it INFN, Laboratori Nucleari di
Frascati, Italy.}\\
$\sharp$ {\it University of California, Los Angeles, USA}

 \vspace{15pt}

$\star${\it Dipartimento di Fisica, Politecnico di Torino \\
C.so Duca degli Abruzzi, 24, I-10129 Torino, and\\
Istituto Nazionale di Fisica Nucleare, Sezione di Torino, \\
Italy}

\vspace{50pt}

{ABSTRACT}

\end{center}

\medskip
We discuss recent results on the interpretation of flux
compactifications on certain Type IIB orientifolds in terms of
gauged ${\mathcal N}$--extended supergravities of no--scale type

\end{titlepage}
\section{Introduction}
Superstring/M--theory are considered to be the most promising
candidates to describe the fundamental theory of gravity. Upon
compactification to four dimensions, the effective low--energy
dynamics of both bulk and brane degrees of freedom is encoded in a
four--dimensional supergravity. Ordinary compactifications
typically yield supergravity models which are far from being
realistic, since they describe a plethora of massless scalar
fields, in part related to the moduli of the internal manifold,
which are not observed in nature and whose v.e.v. define a
continuum of degenerate vacua. In order to derive
phenomenologically viable models from string/M--theory new
dynamics should be introduced, which would be described at the
level of the low--energy effective theory by a suitable scalar
potential $V$. The effect of this potential should be to lift the
vacua degeneracy making the model more predictive and define at
the same time vacua with interesting properties like spontaneous
supersymmetry breaking, cosmological constant etc... Remarkable
progress in this direction has been made in the last four years by
considering compactifications in the presence of non--vanishing
p--form fluxes across cycles of the internal manifold
\cite{frey}-\cite{lrs}. The presence of fluxes determines indeed a
non--trivial scalar potential in the effective low--energy
supergravity, which defines in some cases vacua with vanishing
cosmological constant (at tree level), in which  spontaneous
(partial) supersymmetry breaking may occur and (some of) the
moduli of the internal manifold are fixed. In fact theories with
vanishing cosmological constant are generalized no--scale models,
which were studied long ago in the pure supergravity context
\cite{noscale1,noscale2}. The presence of fluxes gives also rise
in the low--energy supergravity to local symmetries gauged by
vector fields \footnote{In four dimensional supergravities coupled
to linear multiplets, fluxes may give rise to more general
couplings.}. Supergravity models with such gauge symmetries
(gauged supergravities) have been extensively studied in the
literature \cite{dwn}-\cite{dwst1}, also in connection to flux
compactifications or Scherk--Schwarz dimensional reduction
\cite{adfl0}-\cite{vz}.  Actually in extended supergravities
(${\mathcal N}\ge 2$) the \emph{gauging} procedure, which consists
in promoting a global symmetry group of the Lagrangian to local
invariance, is the only way of introducing a non--trivial scalar
potential without explicitly breaking supersymmetry. The global
symmetry group of extended supergravities is the isometry group
$G$ of the scalar manifold, whose non--linear action on the scalar
fields is associated with an electric/magnetic duality action on
the $n_v$ vector field strengths and their duals \cite{gz}. This
duality transformation is required in four dimensions to be
symplectic and thus is defined by the embedding of $G$ inside
${\rm Sp}(2n_v,\mathbb{R})$. Gauge symmetries deriving from flux
compactifications typically are related to non--semisimple Lie
groups ${\Scr G}$ containing abelian translational isometries
acting on axionic fields which originate from ten dimensional R--R
forms $C_{(p)}$ ($p=0,2,4$ for Type IIB) or the NS two form
$B_{(2)}$. The embedding of ${\Scr G}$ inside $G$ is defined at
the level of the corresponding Lie algebras by the flux tensors
themselves, which play the mathematical role of an \emph{embedding
matrix} \cite{dwst1}. \par No--scale models arising from flux
compactifications or Scherk--Schwarz dimensional reduction give
rise to a semi--positive definite scalar potential which has an
interpretation in terms of an ${\mathcal N}$--extended gauged
supergravity in four dimensions. Let us recall the general form of
such scalar potential $V(\Phi)$ ($\Phi$ denoting collectively the
scalar fields)\cite{df}:
\begin{eqnarray}\delta_B^A V(\Phi)=-3 S^{AC}S_{BC}+N^{IA} N_{IB}\,,\label{potential}\end{eqnarray}
where $S_{AB}=S_{BA}$,  and $N^{IA}$ appear in the gravitino and
spin $1/2$ supersymmetry transformations
 \begin{eqnarray} \delta\psi_{A\mu}&=&\frac 1 2
S_{AB}\gamma_\mu\epsilon^B+\cdots\noindent\\
 \delta\lambda^I&=&
N^{IA}\epsilon_A+\cdots\,,\label{varfer}\end{eqnarray} and give
rise  in the supergravity Lagrangian to the following terms:
\begin{eqnarray}\frac{1}{\sqrt{-g}}\mathcal{L}&=&\cdots
+S_{AB}\bar\psi_\mu^A\sigma^{\mu\nu}\psi_\nu^B+iN^{IA}\bar\lambda_I\gamma^\mu\psi_{\mu
A}  -\!V(\Phi)\,.\end{eqnarray}
 Flat space demands that on the extremes ${\partial V}/{\partial\Phi}=0$ the potential  vanishes, so
\begin{eqnarray}3\sum_CS^{AC}S_{CA}&=&\sum_IN^{IA} N_{IA},\qquad \forall A\,,\end{eqnarray}
  The first term in the potential
(\ref{potential}) is the square of the gravitino mass matrix. It
is hermitian, so it can be diagonalized by a unitary
transformation. Assume that it is already diagonal, then the
eigenvalue in the entry $(A_0,A_0)$ is non zero if and only if
$N^{IA_0}\neq 0$ for some $I$. On the other hand, if the gravitino
mass matrix vanishes then $N^{IA}$ must be zero.

 For no-scale models, there is a subset of fields $\lambda^{I'}$ for which
 \begin{eqnarray}3\sum_CS^{AC}S_{CA}&=&\sum_{I'}N^{I'A} N_{I'A},\qquad \forall
A\label{cancel}\end{eqnarray} at any point in the scalar manifold
${\Scr M}_{scal}$. This implies that the potential is given by
 \begin{eqnarray}V(\Phi)&=&\sum_{I\neq I'}N^{IA} N_{IA}\,,\end{eqnarray}
 and it is manifestly positive definite. Zero vacuum energy on a point of ${\Scr M}_{scal}$ implies that
$N^{IA}=0$, $I\neq I'$  at that point. This happens independently
of the number of unbroken supersymmetries, which is controlled by
$N^{I'A}$.\par
 In the sequel we shall
first discuss in some detail the supergravity description of Type
IIB superstring on $K3\times T^2/{\mathbb Z}_2$ orientifold in the
presence of fluxes and D3/D7 branes. Eventually we shall comment
on some general properties of the vacua in no--scale
supergravities originating from flux compactifications and
Scherk--Schwarz dimensional reduction, concluding with a comment
on the dynamical generation of a cosmological constant.
\section{Type IIB on $K3\times T^2/{\mathbb Z}_2$ orientifold with fluxes and D3/D7 branes}
Consider  Type IIB superstring theory compactified on $K3\times
T^2/{\mathbb Z}_2$ orientifold \cite{orientifold1} to four
dimensions \cite{TT}. Let $x^\mu$ ($\mu=0,\dots, 3$) denote the
four dimensional Minkowski coordinates, $x^\ell$
($\ell=4,\dots,7$) the $K3$ coordinates and $x^p$ ($i=8,9$) the
coordinates of $T^2$. The low--energy effective theory is a
${\mathcal N}=2$ supergravity \cite{adfl,adft} which describes the
gravitational multiplet coupled to 3 vector multiplets and 20
hypermultilets. The scalar manifold is the product of a special
K\"ahler manifold spanned by the three complex scalars $s,t,u$ in
the vector multiplets and a quaternionic K\"ahler manifold
describing the 20 hyperscalars\cite{adfl}:
\begin{eqnarray}
\label{scalma} {\Scr M}_{scal}&=& {\Scr M}_{SK}\times {\Scr
M}_{QK}\,,\nonumber\\
  {\Scr M}_{SK}&=&  \left(\frac{{\rm SU} (1,1)}{{\rm U}(1)}\right)_s\times
\left(\frac{{\rm SU}(1,1)}{{\rm U}(1)}\right)_t\times
    \left(\frac{{\rm SU}(1,1)}{{\rm U}(1)}\right)_u\,,\nonumber\\
{\Scr M}_{QK}&=& \frac{{\rm SO}(4,20)}{{\rm SO}(4)\times {\rm
SO}(20)}
\end{eqnarray}
 $s,\,t,\,u$ being complex scalars spanning each factor of $ {\Scr
M}_{SK}$ are defined as follows:
\begin{eqnarray}
s&=& C_{(4)} -{\rm i}\, {\rm Vol} (K_3),\,
\nonumber\\
t&=& \frac{g_{12}}{g_{22}} -{\rm i}\,\frac{\sqrt{{\rm det}
g}}{g_{22}}\,,
\nonumber\\
u&=& C_{(0)} -{\rm i}\, e^{\varphi}\,,
\end{eqnarray}
where $C_{(4)}$ is the axion originating from the components of
the ten dimensional four--form along the directions of $K3$, $
{\rm Vol} (K_3)$ is the volume of $K3$ in the ten dimensional
Einstein frame, $C_{(0)}$ and $\varphi$ are the ten dimensional
axion, dilaton and the matrix $g$ denotes the metric on $T^2$. The
vector fields $A_\mu^\Lambda$ in the bulk sector originate from
the components $B^\alpha_{\mu p}$ of the ten dimensional two forms
$\{B^\alpha_{(2)}\}\equiv \{B_{(2)},\,C_{(2)}\}$ where $\alpha=1,2
$ is the doublet index of the ten dimensional Type IIB duality
group ${\rm SL}(2,\mathbb{R})_u$, and the index $\Lambda=0,\dots,
3$ runs over the ${\bf 4}$ of ${\rm SL}(2,\mathbb{R})_u\times {\rm
SL}(2,\mathbb{R})_t$.\par
 Let
us recall some properties of the $K3$ cohomology. The second order
cohomology group $H^{(2)}(K3,\mathbb{Z})$ is isomorphic to the
lattice $\Gamma^{3,19}$ in which the following inner product
between harmonic two--forms is defined:
$(\alpha,\,\beta)=\int_{K3}\alpha\wedge \beta$. Let us denote by
$\omega_I$, $I=1,\dots, 22$, a basis of $H^{(2)}(K3,\mathbb{Z})$,
and let $m=1,2,3$ and $a=1,\dots, 19$ be the indices running over
the positive and negative signature directions respectively. The
manifold ${\Scr M}_{QK}$ can be written in the form:
\begin{eqnarray}
{\Scr M}_{QK}&=& \left[\frac{{\rm SO}(3,19)}{{\rm SO}(3)\times
{\rm SO}(19)}\times {\rm O}(1,1)\right]\ltimes \{{\bf 22}_{+}\}
\end{eqnarray}
where $ \{{\bf 22}_{+}\}$ denote a subspace generated by 22
abelian isometries $Z_I$ (with positive grading with respect to
the ${\rm O}(1,1) $ generator). These are parametrized by the
axions $C^I$ originating from the components of the four form with
two indices along $K3$ and two indices along $T^2$. The ${\rm
O}(1,1)$ factor is parametrized by the volume of $T^2$:
$\sqrt{{\rm det}(g)}=e^{\phi}$. Finally the 40 complex structure
moduli and the 17 K\"ahler moduli (except the volume) of $K3$ are
described by a $3\times 19$ matrix $e^m{}_a$ which span the ${{\rm
SO}(3,19)}/{{\rm SO}(3)\times {\rm SO}(19)}$ submanifold. These
scalars are arranged in the 20 hyperscalars as follows:
$\{C^m,\,\phi\}$, $\{C^a,\,e^m{}_a\}$.\par Let us now add to the
microscopic setting a stack of $n_3$ space--filling D3 branes and
one of $n_7$ space--filling D7 branes wrapped around $K3$. The
low--energy brane dynamics is described by a SYM theory on their
world volume. We shall consider the SYM theories on the D3/D7
branes to be in the Coulomb phase (namely the branes to be
separated from each other), so that the gauge group and the
massless bosonic modes on the world volume theories are:
\begin{eqnarray}
\mbox{D3:}&&\mbox{gauge group = }\,{\rm
U}(1)^{n_3}\,\,;\,\,\,\mbox{bosonic 0--modes:
}\,A^r_\mu\,\,\,y^r=y^{8,r}+t\,y^{9,r}\,\,(r=1,\dots, n_3)\,,\nonumber\\
\mbox{D7:}&&\mbox{gauge group = }\,{\rm
U}(1)^{n_7}\,\,;\,\,\,\mbox{bosonic 0--modes:
}\,A^k_\mu\,\,\,x^k=x^{8,k}+t\,x^{9,k}\,\,(k=1,\dots,
n_7)\,,\nonumber
\end{eqnarray}
where $y^r$ and $x^k$ are complex scalars describing the position
of each D3, D7--brane along $T^2$ respectively. The massless brane
degrees of freedom will enter the low--energy theory as $n_3+n_7$
additional vector multiplets, causing the special K\"ahler
manifold to enlarge to a homogeneous non--symmetric $3+n_3+n_7$
dimensional  space denoted by $L(0,n_3,n_7)$\cite{dwvp}. The
metric of this manifold was computed in terms of the bulk/brane
fields, using the solvable Lie algebra parametrization, in
\cite{dft}.
\subsection{Geometry of ${\Scr M}_{SK}$}
Let us briefly recall the main formulae of special K\"ahler
geometry. The geometry of the manifold is encoded in the
holomorphic section $\varOmega=(X^\varLambda,\,F_\varSigma)$
which, in the {\it special coordinate} symplectic frame, is
expressed in terms of a prepotential ${\Scr
F}(s,t,u,x^k,y^r)=F(X^\varLambda)/(X^0)^2={\Scr
F}(X^\varLambda/X^0)$, as follows:
\begin{equation}
\varOmega = ( X^\varLambda,\,F_\varLambda=\partial F/\partial
X^\varLambda)\,.\label{specialcoordinate}
\end{equation}
 In our case ${\Scr F}$ is given by
 \begin{equation}
 \label{prepot}
 {\Scr F}(s,t,u,x^k,y^r)\,=\, stu-\frac{1}{2}\,s \,x^k
    x^k-\frac{1}{2}\,u\,
    y^r y^r\,.
\end{equation}
 The
K\"ahler potential $K$ is given by the symplectic invariant
expression:
\begin{equation}
K = -\log \left[{\rm i}(\overline{X}^\varLambda
F_\varLambda-\overline{F}_\varLambda X^\varLambda)\right] \,.
\end{equation}
In terms of $K$ the metric has the form
$g_{i\bar{\jmath}}=\partial_i\partial_{\bar{\jmath}}K$. The
matrices $U^{\varLambda\varSigma}$ and $\overline{{\Scr
N}}_{\varLambda\varSigma}$ are respectively given by:
\begin{eqnarray}
U^{\varLambda\varSigma}&=&e^K\, {\Scr D}_i X^\varLambda {\Scr
D}_{\bar{\jmath}} \overline{X}^\varSigma\,g^{i\bar{\jmath}}=
-\frac{1}{2}\,{\rm Im}({\Scr
N})^{-1}-e^K\,\overline{X}^\varLambda X^\varSigma\,,\nonumber\\
\overline{{\Scr N}}_{\varLambda\varSigma}&=&
\hat{h}_{\varLambda|I}\circ
(\hat{f}^{-1})^I{}_\varSigma\,,\,\,\mbox{where}\,\,\,\,
\hat{f}_{I}^\varLambda = \left(\matrix{{\Scr D}_i X^\varLambda \cr
\overline{X}^\varLambda }\right)\,;\,\,\,\,\hat{h}_{\varLambda|
I}=\left(\matrix{{\Scr D}_i F_\varLambda \cr
\overline{F}_\varLambda }\right) \,.\label{N}
\end{eqnarray}
 For our choice of ${\Scr F}$, $K$ has the following form:\begin{eqnarray}
K &=& -\log [-8\,({\rm Im}(s)\,{\rm Im}(t){\rm
Im}(u)-\frac{1}{2}\,{\rm Im}(s)\,({\rm
Im}(x)^k\,)^2-\nonumber\\&&\frac{1}{2}\,{\rm Im}(u)\,({\rm
Im}(y)^r\,)^2)] \,,
\end{eqnarray}
 with ${\rm Im}(s),\,{\rm Im}(t),\,{\rm
Im}(u)<0$ at $x^k=y^r=0$. The components
$X^\varLambda,\,F_\varSigma$ of the symplectic section which
correctly describe our problem, are chosen by performing a
constant symplectic change of basis from the one in
(\ref{specialcoordinate}) given in terms of the prepotential in
eq. (\ref{prepot}).
 The rotated symplectic sections then become\cite{adft}
\begin{eqnarray}
X^0 &=& \frac{1}{{\sqrt{2}}}\,(1 - t\,u +
\frac{(x^k)^2}{2})\,\,\,\,,\,\,\,\,\, X^1 = -\frac{t +
u}{{\sqrt{2}}}\,,
\nonumber \\
X^2 &=&  -\frac{1}{{\sqrt{2}}}\,({1 + t\,u -
\frac{(x^k)^2}{2}})\,\,\,\,,\,\,\,\,\,X^3 = \frac{t -
u}{{\sqrt{2}}}\,,
\nonumber \\
X^k &=& x^k\,\,\,\,,\,\,\,\,\,X^r = y^r\,,
\nonumber\\
F_0 &=& \frac{s\,\left( 2 - 2\,t\,u + (x^k)^2 \right) +
u\,(y^r)^2}{2\,{\sqrt{2}}}\,\,\,\,,\,\,\,\,\, F_1 =
  \frac{-2\,s\,\left( t + u \right)  +
  (y^r)^2}{2\,{\sqrt{2}}}\nonumber\\
F_2&=&
  \frac{s\,\left( 2 + 2\,t\,u - (x^k)^2 \right)  -
  u\, (y^r)^2}{2\,{\sqrt{2}}}\,\,\,\,,\,\,\,\,\,
 F_3 = \frac{2\,s\,\left( -t + u \right)  +
 (y^r)^2}{{2\,\sqrt{2}}}\nonumber\\
F_i &=& -
  s\,x^k
 \,\,\,\,,\,\,\,\,\,
F_r = -u\,y^r\, .
\end{eqnarray}
Note that, since $\partial X^\varLambda /\partial s =0$ the new
sections do not admit a prepotential, and the no--go theorem on
partial supersymmetry breaking \cite{Cecotti} does not apply in
this case. As in \cite{adfl}, we limit ourselves to gauge
shift--symmetries of the quaternionic manifold of the $K3$
moduli--space. Other gaugings which include the gauge group on the
branes will be considered elsewhere.
\subsection{Fluxes}
Let us consider the effect of switching on fluxes of the
three--form field strengths across cycles of the internal
manifold. The only components of $F^\alpha_{(3)}=dB^\alpha_{(2)}$
which survive the orientifold projection are: $F^\alpha_{(3)}=
F^{\alpha\, I}{}_{p}\, \omega_I\wedge dx^p$. We can describe these
flux components in terms of four integer vectors $f_\Lambda{}^I$,
$\Lambda=0,\dots, 3$ :
\begin{eqnarray}
F^{\alpha\, I}{}_{p}&\equiv &F_\Lambda{}^I=\frac{(4\,\pi^2}{R^3}\,
\alpha^\prime\, f_\Lambda{}^I\,\,\,;\,\,\,\,
f_\Lambda{}^I=\{f_\Lambda{}^m,
h_\Lambda{}^a\}\,\in\,\Gamma^{3,19}\,,
\end{eqnarray}
where  $R$ is the linear size of the internal manifold and last
property follows from the flux quantization condition.\par The
presence of these fluxes imply local invariance in the low--energy
supergravity. A way to see this is to consider the dimensional
reduction of the kinetic term for $C_{(4)}$:
\begin{eqnarray}
D=10&\rightarrow & D=4 \nonumber\\F_{(5)}\wedge
{}^*F_{(5)}&\longrightarrow & (\partial C^I-f_{\Lambda}{}^I\,
A^\Lambda_\mu)^2\,,
\end{eqnarray}
where the four form field strength is defined as:
$F_{(5)}=dC_{(4)}+\frac{1}{2}\,\epsilon_{\alpha\beta}
B_{(2)}^\alpha\wedge F_{(3)}^\beta$. The Stueckelberg--like
kinetic terms for $C^I$ in four dimensions are clearly invariant
under the local translations $C^I\rightarrow C^I+f_\Lambda{}^I
\,\xi^\Lambda$, $\xi^\Lambda$ being four local parameters,
provided the bulk vectors are subject to the gauge transformation
$A^\Lambda_\mu\rightarrow A^\Lambda_\mu+\partial_\mu \xi^\Lambda$.
Thus from general arguments we expect that in the presence of
three form fluxes, the low--energy supergravity should be
invariant under a four dimensional abelian gauge group ${\Scr G}$,
subgroup of $G$ whose generators $X_\Lambda=f_\Lambda{}^I\, Z_I$
are gauged by the bulk vectors. The ${\mathcal N}=2$ supergravity
originated from the flux compactification is obtained therefore
from the ungauged theory through the gauging procedure which
consists in promoting the subgroup ${\Scr G}$ of the isometry
group of ${\Scr M}_{QK}$ to local invariance of the Lagrangian.
Supersymmetry then requires the introduction of additional terms
(fermion shifts) in the fermion supersymmetry transformation
rules, fermion mass terms, and a scalar potential $V(\Phi)$  whose
expression is constrained to be a well defined bilinear in the
fermion shifts \cite{ABCDFFM}. In the sequel we shall denote by
${\Scr P}^x_\Lambda$ and $k^I_\Lambda$ the momentum maps and the
Killing vectors of the gauged isometries $X_\Lambda$:
\begin{eqnarray}
k^I_\Lambda&=&f_\Lambda{}^I\,\,;\,\,\,\,\,{\Scr
P}^x_\Lambda=\sqrt{2}\,e^{\phi}\,([(1+ee^t)^{\frac{1}{2}}]_m{}^x\,
f_\Lambda{}^m+e_a{}^x\, h_\Lambda{}^a)\,.
\end{eqnarray}
In terms of these quantities the scalar potential can be written
as follows:
\begin{eqnarray}
V&=&4\, e^{2\,\phi}\,\left(f^m_\Lambda\,f^m_\Sigma+2\,e^a_m
e^a_n\,f^m_\Lambda\,f^n_\Sigma+h^a_\Lambda\,h^a_\Sigma\right)\,\bar{L}^\Lambda\,
L^\Sigma+\nonumber\\&&
2\,e^{2\,\phi}\,\left(U^{\Lambda\Sigma}-3\,\bar{L}^\Lambda\,
L^\Sigma\right)\,\left(f^m_\Lambda\,f^m_\Sigma+e^a_m
e^a_n\,f^m_\Lambda\,f^n_\Sigma+ 2\,[(1+{\bf e}\,{\bf
e}^T)^{\frac{1}{2}}]^n_m e^n_a\,
f^m_{(\Lambda}\,h^a_{\Sigma)}+\right.\nonumber \\&&\left. e^n_a
e^n_b\,h^a_\Lambda\,h^b_\Sigma\right)\,.\label{pot}
\end{eqnarray}
 Once the potential is known then we can study the vacua of the
theory, that is bosonic backgrounds which extremize $V(\Phi)$. If
we are interested in supersymmetric vacua we need to look for
bosonic backgrounds $\Phi_0$ which admit a Killing spinor
$\epsilon$, namely directions in the supersymmetry parameter space
along which:
\begin{eqnarray}
\delta_\epsilon(\mbox{fermions})_{|\Phi_0}&=&0\,.\end{eqnarray}
 If a
background admits a Killing spinor, it can be shown that it is
also a vacuum of the theory. The spinors of the theory consist of
the gravitini $\psi^A_\mu$, the gaugini $\lambda^{i,A}$
($i=1,\dots, n_v$) and the hyperini $\zeta^{1,A},\,\zeta^{a,A}$.
From the Killing spinor equation $\delta_\epsilon \zeta^{a,A}=0$
we derive the following conditions which should hold for any
supersymmetric vacua:
\begin{eqnarray}
e^a{}_m\, f^m_\Lambda&=&e^m{}_a\,
h^a_\Lambda\,=\,0\,\label{moduli}\\
h^a_\Lambda\, X^\Lambda&=&0\,.\label{tufix}
\end{eqnarray}
Conditions (\ref{moduli}) will fix  $K3$ complex structure moduli,
while eq. (\ref{tufix}) will fix the $T^2$ complex structure $t$
and the axion/dilaton $u$. The Killing spinor equations
$\delta_\epsilon \zeta^{1,A}=0$ and $\delta_\epsilon \psi_\mu^A=0$
turn out to be equivalent for this gauging and, together with the
equations $\delta_\epsilon \lambda^{i,A}=0$, will impose
restrictions on the fluxes.
\paragraph{${\mathcal N}=2$ vacua.}
From the gravitino Killing spinor equation we derive ${\Scr
P}_\Lambda^x\equiv 0$, which, upon implementation of eqs.
(\ref{moduli}) implies
\begin{eqnarray}
f_\Lambda^m &=&0\,,
\end{eqnarray}
which can be restated as the requirement that no flux vector among
the  $f_\Lambda$ in $\Gamma^{3,19}$ have positive norm,
consistently with the results by Tripathy and Trivedi \cite{TT}.
Let us, for the sake of simplicity, choose as the only
non--vanishing components of the flux
\begin{eqnarray}
h_2{}^{a=1}&=&g_2\,\,\,;\,\,\,\,\,h_2{}^{a=2}=g_3\,.
\end{eqnarray}
Condition (\ref{tufix}) then imply:
\begin{eqnarray}
&&X^2=X^3=0\,\,\,\Leftrightarrow\,\,\,t=u \,,\,\,\,
1+t^2=\frac{(x^k)^2}{2}\,,
\end{eqnarray}
so that $t,u$ are fixed, while $s$ and the brane coordinates
$x^k,\,y^r$ remain moduli. Finally conditions (\ref{moduli}) imply
$e^m{}_{a=1,2}=0$. Since the two axions $C^{a=1,2}$ are Goldstone
bosons which provide mass to $A^2_\mu,\,A^3_\mu$, the whole two
hypermultiplets $a=1,2$ will not appear in the low--energy
effective theory. This theory will be no--scale since the
potential at the minimum vanishes identically in the moduli.
\paragraph{${\mathcal N}=1,0$ vacua.} Let us look for ${\mathcal N}=1$ vacua by requiring
 the component
$\epsilon_2$ to be the Killing spinor. Upon implementation of
(\ref{moduli}), we obtain the following conditions:
\begin{eqnarray}
\matrix{\delta_\epsilon \psi_\mu^A=0\cr \delta_\epsilon
\lambda^{i,A}=0}&\Rightarrow
&\cases{(f_\Lambda{}^{x=1}+i\,f_\Lambda{}^{x=2})\, X^\Lambda=0\cr
(f_\Lambda{}^{x=1}+i\,f_\Lambda{}^{x=2})\, \partial_i
X^\Lambda=0\cr f_\Lambda{}^{x=3}=0}\,.\label{psilam}
\end{eqnarray}
Condition $f_\Lambda{}^{x=3}=0$ in particular can be rephrased as
the statement that the flux should be defined by at most two
positive norm vectors in $\Gamma^{3,19}$, consistently with the
\emph{primitivity} condition on the complexified 3--form field
strength $G_{(3)}$ as found by Tripathy and Trivedi \cite{TT}.\par
Suppose, for the sake of simplicity, that the only non--vanishing
flux components are the following
\begin{eqnarray}
f_0{}^{m=1}&=&g_0\,\,\,;\,\,\,\,\,f_1{}^{m=2}=g_1\,\,\,;\,\,\,\,\,h_2{}^{a=1}=g_2\,\,\,;\,\,\,\,\,h_2{}^{a=2}=g_3\,,
\end{eqnarray}
then from the vanishing of the $D7$--brane gaugini variations  in
(\ref{psilam}) we have the condition $x^k=0$, namely that the D7
branes be stuck at the origin of $T^2$. Condition (\ref{tufix})
then implies:
\begin{eqnarray}
&&X^2=X^3=0\,\,\,\Leftrightarrow\,\,\,t=u=-i \,.
\end{eqnarray}
The four axions $C^{m=1,2},\,C^{a=1,2}$ are Goldstone bosons which
provide mass to all the bulk vectors. Finally conditions
(\ref{moduli}) will fix the 40 complex structure moduli of $K3$:
\begin{eqnarray}
e^x{}_{a=1,2}&=&0\,\,\,;\,\,\,\,\,e^{x=1,2}{}_{a>2}=0
\end{eqnarray}
leaving the 17 K\"ahler moduli $e^{x=3}{}_{a>2}$ unfixed. The
unfixed moduli will enter chiral multiplets in the effective
${\mathcal N}=1$ theory as the following complex scalars:
\begin{eqnarray}
s,\,y^r,\,C^{m=3}+i\,e^{\phi},\,C^{a>2}+i\,e^{m=3}{}_{a>2}\,,
\end{eqnarray}
which span the scalar manifold:
\begin{eqnarray}
{\Scr M}_{scal}&=&\frac{{\rm U}(1,1+n_3)}{{\rm U}(1)\times {\rm
U}(1+n_3)}\times \frac{{\rm SO}(2,18)}{{\rm SO}(2)\times {\rm
SO}(18)}\,,
\end{eqnarray}
the former factor being parametrized by $s,\,y^r$. We have not
dealt with all conditions (\ref{psilam}) yet. In particular in the
effective ${\mathcal N}=1$ we can construct a superpotential using
${\mathcal N}=2$ quantities:
\begin{eqnarray}
W&=&[e^{-\phi}\,({\Scr P}^{x=1}_\Lambda+i\,{\Scr
P}^{x=2}_\Lambda)\, X^\Lambda]_{|\Phi_0}\propto
g_0-g_1\,\,\,\mbox{moduli independent}\,.
\end{eqnarray}
On the other hand the expressions in (\ref{psilam})
$(f_\Lambda{}^{x=1}+i\,f_\Lambda{}^{x=2})\, X^\Lambda$ and
$(f_\Lambda{}^{x=1}+i\,f_\Lambda{}^{x=2})\, \partial_i X^\Lambda$
turn all out to be proportional to $g_0-g_1$. Therefore if $W=0$
we have ${\mathcal N}=1$ otherwise the vacuum will break all
supersymmetry. In both cases the potential at the minimum vanishes
identically in the moduli so that the effective supergravity is
no--scale.
\subsection{More general ${\mathcal N}=1$ vacua}
We may generalize the above choice of fluxes so as to have vacua
for more general (complex) values for $t,u$ (in the positivity
domain of the Lagrangian), namely:
\begin{eqnarray}
t&=&
a_t-i\,e^{2\lambda_t}\,\,\,\,;\,\,\,\,\,\,u=a_u-i\,e^{2\lambda_u}\,,\label{moregentu}
\end{eqnarray}
$a_t,\,a_u,\,\lambda_t,\,\lambda_u$ being generic real numbers. To
this end we use the property of the gauged Lagrangian to be still
duality invariant, \emph{provided} we transform under duality
symmetry the fluxes as well. The isometry transformation in ${\rm
SU}(1,1)_t\times {\rm SU}(1,1)_u$ which maps the values $t=u=-i$
into those in (\ref{moregentu}) is represented by the following
symplectic matrix:
\begin{eqnarray}
{\Scr O}&=&{\Scr O}_t\, {\Scr O}_u\,\,\,;\,\,\,\,{\Scr
O}_t=\left(\matrix{A_t^{-1 T}& 0\cr
0&A_t}\right)\,\,\,;\,\,\,\,{\Scr O}_u=\left(\matrix{A_u^{-1 T}&
0\cr C_u& A_u}\right)\,.
\end{eqnarray}
One can verify indeed that
\begin{eqnarray}
{\Scr O}\,\varOmega
(s,t,u,x,y)&=&e^{-\lambda_t-\lambda_u}\varOmega
(s,t^\prime,u^\prime,x^\prime,y^\prime)\,,\nonumber\\
t^\prime&=&
a_t+e^{2\lambda_t}\,t\,\,;\,\,\,u^\prime=a_u+e^{2\lambda_u}\,u\,\,;\,\,\,
y^{\prime r}=e^{\lambda_t}\,y^r\,\,;\,\,\,\, x^{\prime
k}=e^{\lambda_t+\lambda_u}\,x^k\,,
\end{eqnarray}
The flux vectors $f_\Lambda{}^I$  are electric since they fill the
lower part of a symplectic vector. Due to the perturbative form of
${\Scr O}$, its action on the flux vectors will not produce
magnetic charges but will transform them as follows:
\begin{eqnarray}
f_{\Lambda}{}^{\prime m}{}&=&A_{\Lambda}{}^\Sigma\,f_{\Sigma}{}^{
m}{}\,\,\,;\,\,\,\,\,h_{\Lambda}{}^{\prime
a}{}=A_{\Lambda}{}^\Sigma\,h_{\Sigma}{}^{ a}{}\,,\end{eqnarray}
which, in components, read:
\begin{eqnarray}
 f_0{}^{\prime 1}{}&=&\frac{1}{2}\,e^{-{{\lambda }_{ {t}}} - {{\lambda }_{ {u}}}}\,
    \left( 1 + e^{2\,\left( {{\lambda }_{ {t}}} + {{\lambda }_{ {u}}} \right)
           } - {a_{ {t}}}\,{a_{ {u}}} \right)
           \,{g_0}\,\,\,;\,\,\,
f_1{}^{\prime 1}{}=-\frac{1}{2}\, e^{-{{\lambda }_{ {t}}} -
{{\lambda }_{ {u}}}}\,
      \left( {a_{ {t}}} + {a_{ {u}}} \right) \,{g_0}
      \,,\nonumber\\
f_2{}^{\prime 1}{}&=&\frac{1}{2}\,e^{-{{\lambda }_{ {t}}} -
{{\lambda }_{ {u}}}}\,
    \left( 1 - e^{2\,\left( {{\lambda }_{ {t}}} + {{\lambda }_{ {u}}} \right)
           } + {a_{ {t}}}\,{a_{ {u}}} \right)
           \,{g_0}\,\,\,;\,\,\,
f_3{}^{\prime 1}{}=\frac{1}{2}\,e^{-{{\lambda }_{ {t}}} -
{{\lambda }_{ {u}}}}\,
    \left( -{a_{ {t}}} + {a_{ {u}}} \right)
    \,{g_0}\,,\nonumber\\
f_0{}^{\prime 2}{}&=&\left( \frac{e^{-{{\lambda }_{ {t}}} +
{{\lambda }_{ {u}}}}\,
       {a_{ {t}}}}{2} + \frac{e^
        {{{\lambda }_{ {t}}} - {{\lambda }_{ {u}}}}\,{a_{ {u}}}}{2}
    \right) \,{g_1}\,\,\,;\,\,\,
f_1{}^{\prime 2}{}=\left( \frac{e^{{{\lambda }_{ {t}}} - {{\lambda
}_{ {u}}}}}{2} +
    \frac{e^{-{{\lambda }_{ {t}}} + {{\lambda }_{ {u}}}}}{2}
    \right)
    \,{g_1}\,,\nonumber\\
f_2{}^{\prime 2}{}&=&\left( -\frac{ e^{-{{\lambda }_{ {t}}} +
{{\lambda }_{ {u}}}}\,
         {a_{ {t}}}  }{2} -
    \frac{e^{{{\lambda }_{ {t}}} - {{\lambda }_{ {u}}}}\,{a_{ {u}}}}{2}
    \right) \,{g_1}\,\,\,;\,\,\,
f_3{}^{\prime 2}{}=\left( \frac{e^{{{\lambda }_{ {t}}} - {{\lambda
}_{ {u}}}}}{2} -
    \frac{e^{-{{\lambda }_{ {t}}} + {{\lambda }_{ {u}}}}}{2} \right)
    \,{g_1}\,,\nonumber\\
h_0{}^{\prime 1}{}&=&-\frac{1}{2}\, e^{-{{\lambda }_{ {t}}} -
{{\lambda }_{ {u}}}}\,
      \left( -1 + e^{2\,\left( {{\lambda }_{ {t}}} + {{\lambda }_{ {u}}} \right)
             } + {a_{ {t}}}\,{a_{ {u}}} \right) \,{g_2}
             \,\,\,;\,\,\,
h_1{}^{\prime 1}{}=-\frac{1}{2}\, e^{-{{\lambda }_{ {t}}} -
{{\lambda }_{ {u}}}}\,
      \left( {a_{ {t}}} + {a_{ {u}}} \right) \,{g_2}
    \,,\nonumber\\
h_2{}^{\prime 1}{}&=&\frac{1}{2}\,e^{-{{\lambda }_{ {t}}} -
{{\lambda }_{ {u}}}}\,
    \left( 1 + e^{2\,\left( {{\lambda }_{ {t}}} + {{\lambda }_{ {u}}} \right)
           } + {a_{ {t}}}\,{a_{ {u}}} \right)
           \,{g_2}\,\,\,;\,\,\,
h_3{}^{\prime 1}{}=\frac{1}{2}\,e^{-{{\lambda }_{ {t}}} -
{{\lambda }_{ {u}}}}\,
    \left( -{a_{ {t}}} + {a_{ {u}}} \right)
    \,{g_2}\,,\nonumber\\
h_0{}^{\prime 2}{}&=&\left( -\frac{e^{-{{\lambda }_{ {t}}} +
{{\lambda }_{ {u}}}}\,
         {a_{ {t}}}  }{2} +
    \frac{e^{{{\lambda }_{ {t}}} - {{\lambda }_{ {u}}}}\,{a_{ {u}}}}{2}
    \right) \,{g_3}\,\,\,;\,\,\,
h_1{}^{\prime 2}{}=\left( \frac{e^{{{\lambda }_{ {t}}} - {{\lambda
}_{ {u}}}}}{2} -
    \frac{e^{-{{\lambda }_{ {t}}} + {{\lambda }_{ {u}}}}}{2} \right) \,{g_3}
    \,,\nonumber\\
h_2{}^{\prime 2}{}&=&\left( \frac{e^{-{{\lambda }_{ {t}}} +
{{\lambda }_{ {u}}}}\,
       {a_{ {t}}}}{2} - \frac{e^
        {{{\lambda }_{ {t}}} - {{\lambda }_{ {u}}}}\,{a_{ {u}}}}{2}
    \right) \,{g_3}\,\,\,;\,\,\,
h_3{}^{\prime 2}{}=\left( \frac{e^{{{\lambda }_{ {t}}} - {{\lambda
}_{ {u}}}}}{2} +
    \frac{e^{-{{\lambda }_{ {t}}} + {{\lambda }_{ {u}}}}}{2} \right)
    \,{g_3}\label{flux2}
\end{eqnarray}
One can verify that with this choice of fluxes ${\mathcal N}=1$
residual supersymmetry imply $g_0=g_1$, $x^k=0$ and $t,u$ fixed at
the values in (\ref{moregentu}).\par The possibility of fixing the
effective string coupling constant to small values as important
implications. For instance it makes it possible to apply the model
to the construction of inflationary models
\cite{hybrid}-\cite{KTW}, in which the \emph{slow-roll} of the
inflaton (one of the $y^r$ moduli) is realized  once perturbative
corrections to the potential are taken into account\footnote{We
are grateful to R. Kallosh for explaining this point to us.}.
\paragraph{ D7 brane world volume fluxes.} Within this
framework we can consider the effect of switching on fluxes of the
D7 gauge field strengths ${\mathcal F}^k_{\mu\nu}$ across two
cycles of $K3$. This corresponds for instance to gauging
additional $Z^a$ isometries by means of D7 brane vectors $A^k_\mu$
\cite{adft} . The constant Killing vectors are
$k^u_\Lambda=g^k_4$, $\Lambda=3+k,\, k=1,\dots,n_7$, along the
direction $q^u=C^{a=3,\dots,2+n_7}$ (recall that the isometries
$Z_{a=1,2}$ have already been gauged by the vectors
$A^{2,3}_\mu$).\par As far as supersymmetric vacua are concerned,
from inspection of the fermion shifts
 it is straightforward to verify that the existence of a
 constant Killing spinor always requires $X^{2},\,X^3,\,X^{3+k}=0$
 which implies $x^k=0$ and $t=u=-{\rm i}$ even in the ${\mathcal N}=2$ case
 (still corresponding to the choice $g_0=g_1=0$).
  As before we have
 ${\mathcal N}=1$ if $g_0=g_1\neq 0$ and ${\mathcal N}=0$ otherwise.
 \section{No--scale supergravity from  Scherk--Schwarz generalized dimensional reduction.}
 As pointed out in the introduction, spontaneously broken supergravity can also be obtained through a
 Scherk--Schwarz dimensional reduction from $D+1$ to $D$
 dimensions \cite{ss,adfl0,dwst1,bglnor}. In order for the theory to admit a stable vacuum the
 Scherk--Schwarz phases should be taken to be in the Cartan
 subalgebra of the maximal compact subgroup of the isometry group
 $G$ of the theory in $D+1$ dimensions. The scalar potential is
 obtained from the non--linear $\sigma$--model describing the $D+1$
 dimensional scalar fields:
 \begin{eqnarray}
\sqrt{-{\rm det}(g)}\, g^{\mu\nu}\,P_\mu^I\, P^{I}_\nu\,,
 \end{eqnarray}
where $P^I_\mu$ are the pull--back on space--time of the vielbeins
$P^I_i$ of the $D+1$ dimensional scalar manifold. By taking
$\mu=\nu=D+1$ we have the following potential in $D$ dimensions:
 \begin{eqnarray}
V&=& e^{-2\,\frac{D-1}{D-2}\sigma}\,P_{D+1}^I\, P^{I}_{D+1}\,\ge
\,0\,,
 \end{eqnarray}
where $\sigma$ is the modulus associated to the radius of the
internal dimension and $P_{D+1}^I=P_i^I\, M_j{}^i\, \phi^j$,
$M_i{}^j$ being the Scherk--Schwarz phases. The potential has an
absolute minimum (at the origin of the scalar manifold) only if
$M_i{}^j$ besides being a global symmetry of the $D+1$ theory is
also compact, so that there exist a point in the moduli space in
which  $P_{D+1}^I=0$. All the scalars are fixed at this minimum
except $\sigma $ and all the $D+1$ dimensional scalars $\phi^i$
for which $M_i{}^j\, \phi^j=0$. Finally the gravitino mass matrix
is provided by $Q_{D+1}$ which is the pull--back on the direction
$D+1$ of the R--symmetry connection $Q_i$ on the scalar manifold
in  $D+1$ dimensions.

 \section{Type IIB on $T^6/\mathbb{Z}_2$ orientifold with fluxes and D3 branes}
 As a final example of let us briefly mention
  the gauged supergravity which describes the (classical) low--energy
limit of Type IIB on $T_6/\mathbb{Z}_2$ orientifold in the
presence of space--filling D3 branes and three form NS and RR
fluxes \cite{FP,dfv,DFGVT,bhk}. It is an ${\mathcal N}=4$ model
with an abelian gauge symmetry generated by twelve independent
combinations of the fifteen translational isometries acting on the
axions which originate from the internal components of the ten
dimensional 4--form $C_{(4)}$. This model exhibits vacua with
vanishing cosmological constant at tree level  and a hierarchical
supersymmetry breaking ${\mathcal N}=4\rightarrow 3\rightarrow 2
\rightarrow 1\rightarrow 0$ in which the masses of the gravitini
are provided by four independent flux parameters $m_i$,
$i=1,\dots, 4$, expressed in units of $\alpha^\prime/{\rm
Vol}(T^6)^{\frac{1}{2}}$.
\section{No--scale supergravities and the cosmological constant}
All the models discussed above exhibit partial super--Higgs around
Minkowski vacua. Let us comment on the one--loop corrections to
the cosmological constant. We start recalling that the quartic,
quadratic and logarithmic divergent parts, in any field theory,
are respectively controlled by the following coefficients
\begin{eqnarray}
{\rm Str}({\Scr M}^{2k})&=&
\sum_{J}(-)^{2J}\,(2J+1)\,m_J^{2k}\,\,;\,\,\,\ k=0,1,2\,.
\end{eqnarray}
On the other hand, the sum rules
\begin{eqnarray}
{\rm Str}({\Scr M}^{2k})&=& 0\,\,;\,\,\,\ k<{\mathcal N}\,,
\end{eqnarray}
in ${\mathcal N}$--extended supergravity seem to be of general
validity for theories where a ${\mathcal N}\rightarrow {\mathcal
N}-1$ breaking is possible \cite{adfl2}. This requires long
massive gravitino multiplets since the massive gravitino is
Majorana and therefore cannot be BPS. On the other hand, for
theories with central charges, like the Scherk--Schwarz breaking
of ${\mathcal N}=8$, gravitini are pairwise degenerate and the
same sum rules apply only for $k<{\mathcal N}/2$, ${\mathcal N}$
being even. It is important to note that the bulk sector of
${\mathcal N}=4$ Type IIB orientifold with fluxes does indeed
coincide with a $\mathbb{Z}_2$ truncation of the ${\mathcal N}=8$
Scherk--Schwarz supergravity, as was shown in \cite{DFGVT}.
Similarly ${\mathcal N}\le 6$  Scherk--Schwarz supergravities, by
$\mathbb{Z}_2$ reduction which removes the gravitino degeneracy,
satisfy the same sum rules as the parent theory \cite{vz}.
\par As an example
let us consider the ${\mathcal N}=4$ no--scale model from Type IIB
on $T^6/\mathbb{Z}_2$ orientifold. In this case it was shown that
${\rm Str}({\Scr M}^2)={\rm Str}({\Scr M}^4)={\rm Str}({\Scr
M}^6)=0$ while from general arguments one would expect ${\rm
Str}({\Scr M}^8)\propto m_1^2 m_2^2 m_3^2 m_4^2\neq 0$. The first
finite contribution to the cosmological constant would then be:
\begin{eqnarray}
\Lambda&\sim & \frac{m_1^2 m_2^2 m_3^2 m_4^2}{M_{Pl}^4}
\end{eqnarray}
It is intriguing to note that, if the supersymmetry breaking scale
is taken to be $m_1\sim m_2\sim m_3\sim m_4\sim 10\, TeV\sim
10^{-15} M_{Pl}$ then we would obtain from the above formula
$\Lambda\sim 10^{-120} M_{Pl}^4$ which is consistent with the most
recent experimental data \cite{experiment}.



\section*{Acknowledgements}
This report is mainly based on work done in collaboration with L.
Andrianopoli, M. Lled\'o and C. Angelantonj. M.T. and S.F. would
like to thank R. Kallosh, G. Villadoro and Zwirner for useful
discussions.
 Work
supported in part by the European Community's Human Potential
Program under contract HPRN-CT-2000-00131 Quantum Space-Time, in
which  R. D'A. is associated to Torino University. The work of
S.F. has been supported in part by European Community's Human
Potential Program under contract HPRN-CT-2000-00131 Quantum
Space-Time, in association with INFN Frascati National
Laboratories and by D.O.E. grant DE-FG03-91ER40662, Task C.

\end{document}